\documentclass[conference]{IEEEtran}
\IEEEoverridecommandlockouts
\usepackage{cite}
\usepackage{amsmath,amssymb,amsfonts}
\usepackage{algorithm}
\usepackage{algorithmic}
\usepackage{graphicx}
\usepackage{textcomp}
\usepackage{xcolor}

\usepackage{graphicx}
\usepackage{caption}
\usepackage{paralist}
\usepackage{amssymb}
\usepackage{amsmath}
\usepackage{amsfonts}
\usepackage{tabularx}
\usepackage[
  paper=letterpaper,
  top=0.77in,       
  bottom=1.1in,       
  left=0.635in,     
  right=0.635in,
]{geometry}
\usepackage{booktabs}
\usepackage{array}

\def\BibTeX{{\rm B\kern-.05em{\sc i\kern-.025em b}\kern-.08em
  T\kern-.1667em\lower.7ex\hbox{E}\kern-.125emX}}

\makeatletter
\providecommand{\linebreakand}{%
  \end{@IEEEauthorhalign}%
  \hfill\mbox{}\par
  \mbox{}\hfill\begin{@IEEEauthorhalign}%
}
\makeatother
\begin{document}


\author{\IEEEauthorblockN{Mayank Raj}
\IEEEauthorblockA{
\textit{University of Massachusetts Dartmouth}\\
Dartmouth, MA USA\\
mraj1@umassd.edu}
\and
\IEEEauthorblockN{Nathaniel D. Bastian}
\IEEEauthorblockA{\textit{United States Military Academy}\\
West Point, NY USA \\
nathaniel.bastian@westpoint.edu}
\and
\IEEEauthorblockN{Lance Fiondella}
\IEEEauthorblockA{\textit{University of Massachusetts Dartmouth}\\
Dartmouth, MA USA \\
lfiondella@umassd.edu}
\linebreakand
\IEEEauthorblockN{G\"{o}khan Kul}
\IEEEauthorblockA{\textit{University of Massachusetts Dartmouth}\\
Dartmouth, MA USA \\
gkul@umassd.edu}
}

\title{Synthetic Network Packet Generation through Statistical Learning and Genetic Algorithms}

\maketitle

\begin{abstract}
Developing robust intrusion detection systems (IDS) for Internet of Things (IoT) environments requires large, labeled datasets capturing realistic traffic distributions across both benign and malicious activity. Existing public datasets suffer from fixed activity distributions and extreme class imbalance, while deep generative models (GANs, VAEs) provide no mechanism to enforce that synthetic packets remain within physically valid feature ranges. This paper proposes and compares two constraint-enforcing approaches for synthetic IoT network packet generation: (i)~a statistical learning method combining PCA-based latent space sampling with dual One-Class SVM (OCSVM) and Isolation Forest (IF) boundary enforcement, and (ii)~a genetic algorithm (GA) method that treats packet generation as a multi-objective optimization problem with explicit fitness criteria for anomaly model acceptance and distributional fidelity. Both methods embed hard validity constraints, namely dual anomaly-detection gating, feature-range clamping, and independent validation, directly into the synthesis pipeline. Evaluation on the complete ACI IoT 2023 dataset (1,231,411 packets, 12 attack categories, class imbalance up to 175,805:1) demonstrates that both methods achieve PASS status across all categories under independently trained validators with a 30\% anomaly rate threshold: the statistical method attains 1.20\% average anomaly rate with $\sim$1,091 packets/s throughput, while the GA attains 0.62\% average anomaly rate with organic per-class variance (0.00\%--2.50\%) at $\sim$5.7 packets/s. Both methods successfully amplify the 5-sample ARP Spoofing category by $200\times$ to 1,000 validated packets. The $\sim$190:1 throughput ratio between methods, combined with their complementary quality profiles, provides evidence-based selection criteria for deployment contexts ranging from rapid dataset augmentation to adversarial robustness testing.
\end{abstract}


\section{Introduction}

The increasing number of Internet of Things (IoT) devices in critical systems, government networks, and military infrastructures has increased the attack surface that adversaries can target~\cite{TAHSIEN2020102630}. Developing effective intrusion detection systems (IDS) for these environments requires large, labeled datasets that capture realistic traffic distributions across both benign and malicious activity. However, constructing such datasets demands extensive virtualization, coordinated attack simulation, and expert-driven labeling, all of which remain scarce in practice~\cite{7307098}.

Many existing public datasets do not exhibit the device-constrained traffic patterns and protocol heterogeneity that characterize contemporary IoT deployments~\cite{7307098}. IoT-specific traces with contemporary attack scenarios are provided by more recent initiatives such as ACI IoT 2023~\cite{10773916} and Bot-IoT~\cite{KORONIOTIS2019779}. However, their fixed activity distributions limit their applicability for evaluating IDS robustness under various operating conditions.

Synthetic data generation offers a path forward, but existing approaches present critical limitations. When applied to IoT traffic, deep generative models like GANs and VAEs experience mode collapse and unstable convergence~\cite{10.1145/3600160.3605031, 9448103}. Additionally, they lack a way to ensure that produced packets stay within physically valid feature ranges~\cite{9671580}. Adversarial perturbation techniques like FGSM~\cite{naqvi2023adversarial, zhang2023generate} and PGD~\cite{9191288} further exacerbate this issue by generating samples that defy the structural restrictions defining legitimate network traffic.

This enforcement gap raises an important question: can we create generation pipelines that incorporate \textit{hard validity constraints}, namely explicit boundary enforcement and anomaly-model gating, directly into the synthesis process instead of depending solely on implicit distributional learning? If so, what generation paradigm better serves this purpose, and under what data availability conditions?

To investigate these questions systematically, this paper proposes and compares two constraint-enforcing approaches for generating synthetic IoT network packets: (i)~a \textit{statistical learning} method combining PCA-based latent space sampling with One-Class SVM (OCSVM) and Isolation Forest (IF) boundary enforcement, and (ii)~a \textit{genetic algorithm} (GA) method that treats packet generation as a multi-objective optimization problem with explicit fitness criteria for anomaly model acceptance and distributional fidelity. Both methods share a critical architectural feature: a dual anomaly-detection gate that requires concurrent acceptance by independently trained OCSVM and IF models before any synthetic packet is admitted to the output set. This stands in contrast to GAN-based methods, where quality control is an emergent property of the adversarial training process rather than an enforceable constraint.

We formalize the generation objective as follows. Given a real dataset $\mathcal{D} = \{(\mathbf{x}_i, y_i)\}_{i=1}^{N}$ with $\mathbf{x}_i \in \mathbb{R}^d$ and $y_i \in \{c_1, \ldots, c_K\}$, we seek to produce, for each class $c_k$, a synthetic set $\mathcal{S}_k = \{\hat{\mathbf{x}}_j\}_{j=1}^{M}$ satisfying four criteria:
\begin{enumerate}
    \item Each $\hat{\mathbf{x}}_j$ must pass concurrent OCSVM and IF acceptance ($f_{\text{OCSVM}}^{(\text{gen})}(\hat{\mathbf{x}}_j) = +1 \;\wedge\; f_{\text{IF}}^{(\text{gen})}(\hat{\mathbf{x}}_j) = +1$) during synthesis; failing candidates are discarded and regenerated
    \item $\hat{x}_{j,i} \in [\min(\mathcal{D}_{k,i}),\, \max(\mathcal{D}_{k,i})]$ for all features $i$, enforced via element-wise clamping for boundary compliance
    \item The anomaly rate $A_f = \frac{1}{M} \sum_{j=1}^{M} \mathbf{1}[f^{(\text{val})}(\hat{\mathbf{x}}_j) = -1] < \tau = 0.30$ when evaluated by separately trained models $f^{(\text{val})}$ sharing no parameters with $f^{(\text{gen})}$, for independent validation
    \item Synthetic sets are produced for all $K$ categories, including those with extreme scarcity ($n_k \ll 100$), without degradation below $\tau$
\end{enumerate}

We evaluate both methods on the complete ACI IoT 2023 dataset~\cite{10773916}, comprising 1,231,411 labeled packets across 12 attack categories with class imbalance ratios reaching 175,805:1 (Benign $n{=}879{,}027$ vs.\ ARP Spoofing $n{=}5$). This extreme imbalance mirrors operational reality, where novel attack vectors are precisely the categories for which defenders most need training data yet have the fewest samples.

Specifically, this work addresses three research questions:

\smallskip
\noindent\textbf{RQ1:} \textit{Does embedding dual anomaly-detection gating and feature-range clamping directly into the generation pipeline produce synthetic IoT packets that independently trained validators consistently accept as normal-for-class?} Existing generators offer no mechanism to reject individual samples violating physical feature constraints~\cite{10.1145/3600160.3605031, 9448103}.

\smallskip
\noindent\textbf{RQ2:} \textit{To what extent can constraint-enforcing generation amplify extremely scarce attack categories, down to as few as five real samples, while maintaining anomaly rates below the acceptance threshold?} Standard oversampling (SMOTE~\cite{chawla2002smote}) assumes sufficient neighborhood density, and deep generative models require substantial samples; both assumptions fail at $n{=}5$~\cite{10.1145/3689935.3690396, 10765948}.

\smallskip
\noindent\textbf{RQ3:} \textit{What are the quantitative trade-offs between single-pass statistical sampling and iterative evolutionary optimization for constraint-enforcing IoT packet generation?} We quantify anomaly rates, generation throughput, and per-class variance to yield evidence-based method selection criteria~\cite{9807332}.

\smallskip
\noindent The contributions of this paper are:
\begin{enumerate}
    \item A systematic comparison of statistical learning and GA approaches for constraint-enforcing synthetic IoT packet generation, with four explicit acceptance criteria (Section~\ref{sec:metrics}).
    \item An independent validation framework using separately trained OCSVM and IF models at global and class-specific granularity, ensuring no shared parameters between generation and assessment (Section~\ref{sec:validation}).
    \item Empirical evidence that both methods achieve PASS status across all 12 ACI IoT 2023 categories under $\tau = 0.30$: the statistical method attains 1.20\% average anomaly rate; the GA attains 0.62\%, including $200\times$ amplification of the 5-sample ARP Spoofing category (Section~\ref{sec:results}).
    \item A computational trade-off analysis showing ${\sim}190{:}1$ throughput ratio between the two methods, with deployment-context selection guidelines (Section~\ref{sec:computational}).
\end{enumerate}


\section{Background and Related Work}

\subsection{IoT Network Traffic Datasets}

The variety, scale, and temporal dynamics of IoT environments are not captured by the packet distributions of enterprise-centric corpora, which have historically been the norm of network security research~\cite{7307098}. While CICIDS2017~\cite{sharafaldin2018cicids} and UNSW-NB15~\cite{moustafa2015unsw} incorporated modern attack profiles, they are still enterprise-focused and do not include IoT-specific protocol characteristics (MQTT, CoAP, Zigbee, BLE). 
The Bot-IoT dataset~\cite{KORONIOTIS2019779} advanced IoT-specific data collection by capturing large-scale botnet activity, but remained confined to a single attack family. The ACI IoT 2023 dataset~\cite{10773916} addresses this limitation with over 1.2 million labeled packets from wired and wireless IoT testbeds across twelve attack categories, substantially reducing the gap between experimental and field deployments. Nevertheless, its fixed activity distributions and extreme class imbalance (five orders of magnitude between largest and smallest classes) limit both generation and detection research without augmentation.

\subsection{Synthetic Data Generation for Network Security}

Synthetic generation has emerged as a pragmatic alternative when empirical traces are scarce or imbalanced. GANs learn to produce samples through adversarial training, while VAEs sample from learned latent representations~\cite{10.1145/3600160.3605031}. Conditional variants such as CTGAN~\cite{xu2019modeling} extend this to class-specific tabular synthesis. However, when applied to IoT traffic, these models face mode collapse, class-conditioned drift, and unstable training due to heterogeneous feature types (binary flags, continuous counters, categorical fields)~\cite{9448103}.

Traditional oversampling via SMOTE~\cite{chawla2002smote} interpolates between nearest neighbors and has been widely applied to IDS class imbalance~\cite{FERRAG2020102419}. However, SMOTE assumes sufficient local density; at $n{=}5$ samples, the neighborhood structure becomes degenerate, producing physically implausible feature combinations.

Genetic Algorithms have been applied to cybersecurity for feature selection and multi-objective optimization~\cite{FERRAG2020102419}, but no prior work has formulated IoT packet generation as a constrained evolutionary problem with dual anomaly-model fitness evaluation.

Recent work by Liu et al.~\cite{liu2025synthetic} evaluated three generative AI methods: CTGAN, TabDDPM, and GReaT (LLM-based), for synthetic network flow generation across UNSW-NB15 ~\cite{moustafa2015unsw}, CICIDS2017 ~\cite{sharafaldin2018cicids}, and CICDDoS2019 datasets ~\cite{sharafaldin2019cicddos}. Their results demonstrate that LLM-based generation achieves up to 86\% macro-average F1 when training downstream classifiers on synthetic data, outperforming distribution-based alternatives on imbalanced datasets. However, none of the three methods incorporate explicit validity constraints during generation. Synthetic packets are evaluated only post-hoc via downstream classifier performance, with no mechanism to reject individual packets that violate physical feature bounds or fail anomaly-model acceptance during synthesis. Our work addresses this limitation by embedding dual OCSVM and Isolation Forest gates directly into the generation pipeline, ensuring that every synthetic packet satisfies hard validity criteria before admission to the output set.

A critical gap across all existing methods is the absence of explicit validity enforcement. Our work addresses this through constraint-enforcing generation, where dual OCSVM and Isolation Forest gates serve as hard acceptance criteria during synthesis.

\subsection{Adversarial Robustness of Network Intrusion Detectors}

It is commonly known that ML-based IDS are susceptible to adversarial manipulation~\cite{7467366, 7958570}. Hashemi and Keller~\cite{9289869} showed how to evade NIDS adversarially in the network security domain, while gradient-based attacks (FGSM~\cite{naqvi2023adversarial, zhang2023generate}, PGD~\cite{9191288}) significantly impair performance~\cite{9671580}. Robustness is addressed in recent work using CNN reliability assessment~\cite{10765948}, XAI~\cite{9807332}, zero-day recognition~\cite{10773707}, and meta-learning~\cite{10.1145/3689935.3690396}. These results demonstrate that training data diversity plays a crucial role in IDS robustness, which drives our constraint-enforcing strategy for producing legitimate synthetic traffic across underrepresented threat categories.

\section{Methodology}
\label{sec:methodology}

\subsection{System Architecture}

\begin{figure}[t]
    \centering
    \includegraphics[width=\columnwidth]{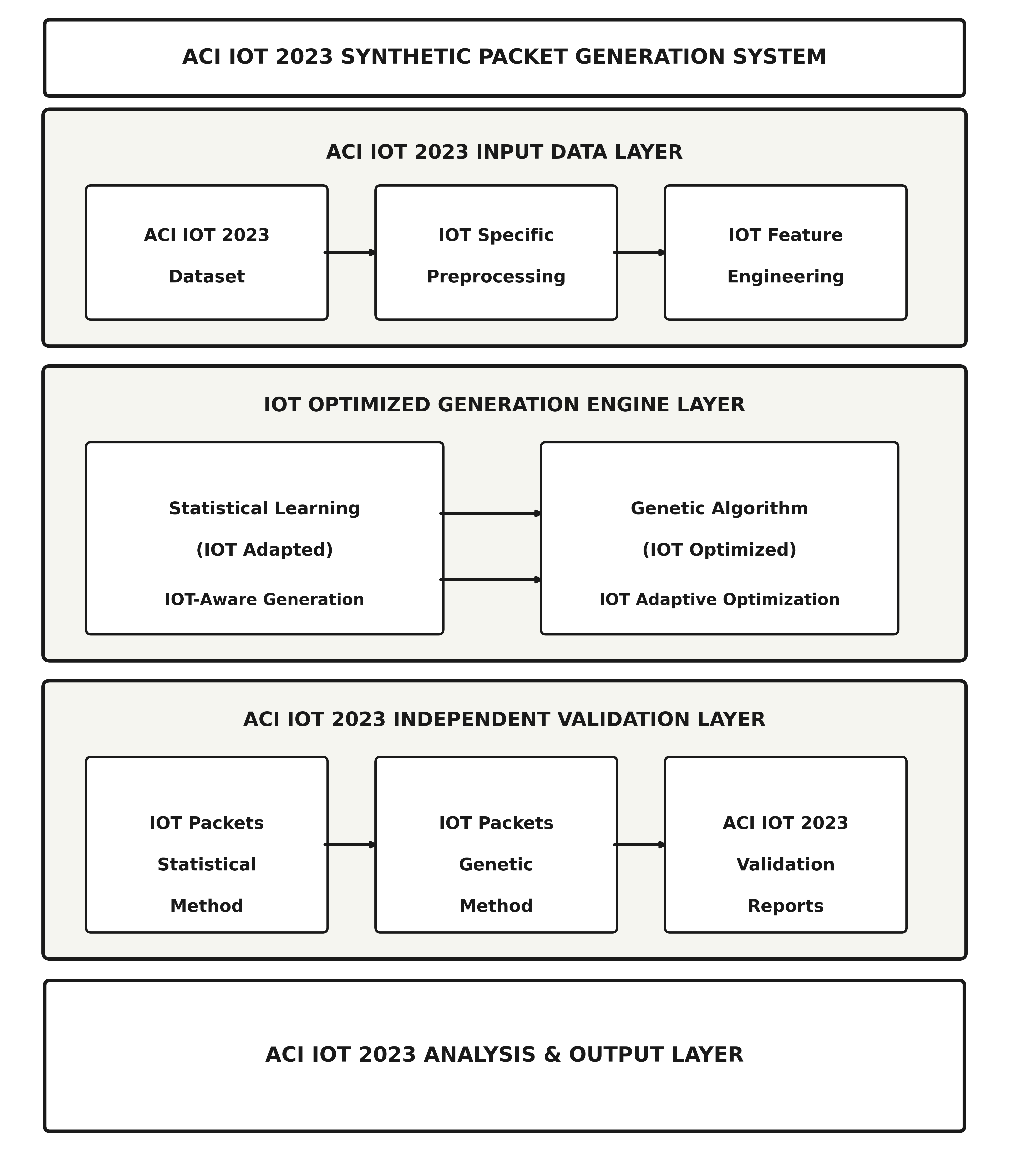}
    \caption{Four-layer system architecture. Layer 1 extracts all dataset-derived parameters from ACI IoT 2023: $N{=}1{,}231{,}411$ packets, $d{=}75$ active features after preprocessing, per-class statistics $\boldsymbol{\mu}_c$, $\boldsymbol{\sigma}_c$, $\mathbf{l}_c$, $\mathbf{u}_c$, and class labels $y_i \in \{c_1, \ldots, c_{12}\}$. Layer 2 applies design parameters ($\alpha{=}0.7$, $P{=}200$, $p_m{=}0.05$) to two parallel generation pipelines with dual OCSVM+IF gating. Layer 3 validates outputs using independently trained models ($\tau{=}0.30$) with no shared parameters. Layer 4 computes quality metrics (Eqs.~\ref{eq:anomaly_rate}--\ref{eq:fidelity}).}
    \label{fig:architecture}
\end{figure}

Figure~\ref{fig:architecture} illustrates the four-layer architecture. The design isolates generation from validation to ensure objective quality assessment.

\textbf{Layer 1: Input Data.} Obtains the raw ACI IoT 2023 CSV containing 85 features and 1,231,411 packets. Preprocessing reduces active features from 85 to 75 by removing constant-variance columns and using feature-wise imputation to fill in missing flow-rate values. \texttt{StandardScaler} normalization guarantees consistent feature scales across cumulative counters and binary flags. This preprocessed ACI IoT 2023 data is the only source of parameters needed in later layers.

\textbf{Layer 2: Generation Engine.} Executes both generation methods operating on identical preprocessed data: (1) the statistical learning pipeline (Section~\ref{sec:stat}) and (2) the genetic algorithm pipeline (Section~\ref{sec:ga}). Both methods train class-specific OCSVM and Isolation Forest models on the preprocessed features and use these models as dual constraint enforcers during generation.

\textbf{Layer 3: Independent Validation.} Employs separately trained OCSVM and Isolation Forest models on original data to assess synthetic packet quality without information leakage from the generation phase (Section~\ref{sec:validation}).

\textbf{Layer 4: Analysis \& Output.} Computes the formal quality metrics defined in Section~\ref{sec:metrics} and produces per-class and aggregate comparative reports.

\subsection{Quality Metrics}
\label{sec:metrics}

We define three quantitative metrics to evaluate synthetic packet quality:

\textbf{Definition 1 (Anomaly Rate).} For a set of $M$ synthetic packets $\{\hat{x}_1, \dots, \hat{x}_M\}$ evaluated by an anomaly detector $f$:
\begin{equation}
    A_f = \frac{1}{M} \sum_{j=1}^{M} \mathbf{1}[f(\hat{x}_j) = -1]
\label{eq:anomaly_rate}
\end{equation}
where $f(\hat{x}_j) \in \{-1, +1\}$ denotes the anomaly or normal prediction, respectively. A synthetic packet set achieves PASS status if $A_f < \tau$ where $\tau = 0.30$.

\textbf{Definition 2 (Boundary Compliance Rate).} For class $c$ with observed feature bounds $[\mathbf{l}_c, \mathbf{u}_c] \in \mathbb{R}^d$:
\begin{equation}
    B_c = \frac{1}{M} \sum_{j=1}^{M} \left[\mathbf{l}_c \leq \hat{x}_j \leq \mathbf{u}_c\right]
\label{eq:boundary}
\end{equation}
where the inequality is element-wise across all $d = 75$ features. Both generation methods enforce $B_c = 1.0$ by construction through element-wise clamping (\texttt{np.clip}) to $[\mathbf{l}_c, \mathbf{u}_c]$ prior to acceptance testing.

\textbf{Definition 3 (Distributional Fidelity).} The normalized Euclidean distance between a synthetic packet $\hat{x}_j$ and the class centroid $\boldsymbol{\mu}_c$ with standard deviation $\boldsymbol{\sigma}_c$:
\begin{equation}
    D(\hat{x}_j, c) = \frac{1}{\sqrt{d}} \left\| \frac{\hat{x}_j - \boldsymbol{\mu}_c}{\boldsymbol{\sigma}_c} \right\|_2
\label{eq:fidelity}
\end{equation}
Lower values indicate closer adherence to the learned class distribution. This metric is used directly in the GA fitness function (Eq.~\eqref{eq:fitness}) and serves as a diagnostic for the statistical learning method.

\subsection{Data Preprocessing}
\label{sec:preprocess}

Table~\ref{tab:dataset} summarizes the ACI IoT 2023 class distribution. Feature-wise imputation filled 3,848 missing values (0.16\%) in \texttt{Flow Bytes/s} and \texttt{Flow Packets/s}, which are fields commonly absent in zero-duration IoT flows. Six constant-variance columns (\texttt{Bwd PSH Flags}, \texttt{Bwd URG Flags}, \texttt{Fwd Bytes/Bulk Avg}, \texttt{Fwd Packet/Bulk Avg}, \texttt{Fwd Bulk Rate Avg}, \texttt{Subflow Bwd Packets}) were removed, and the categorical \texttt{Connection Type} feature was one-hot encoded into binary wired/wireless indicators (60\%/40\% split in the dataset), reducing the active feature set from 85 to $d = 75$.

\begin{table}[t]
\centering
\caption{ACI IoT 2023 Dataset Class Distribution}
\label{tab:dataset}
\begin{tabular}{@{}lrrc@{}}
\toprule
\textbf{Attack Category} & \textbf{Count} & \textbf{\%} & \textbf{Imbalance Ratio} \\
\midrule
Benign          & 879,027 & 71.36 & 1 : 1 (baseline)\\
Port Scan       & 179,130 & 14.55 & 4.9 : 1 \\
ICMP Flood      &  91,226 &  7.41 & 9.6 : 1 \\
Ping Sweep      &  29,433 &  2.39 & 29.9 : 1 \\
DNS Flood       &  19,069 &  1.55 & 46.1 : 1 \\
Vulnerability Scan & 15,915 & 1.29 & 55.2 : 1 \\
OS Scan         &  15,399 &  1.25 & 57.1 : 1 \\
Dictionary Attack & 12,673 & 1.03 & 69.4 : 1 \\
Slowloris       &   7,628 &  0.62 & 115.2 : 1 \\
UDP Flood       &   6,871 &  0.56 & 127.9 : 1 \\
SYN Flood       &   5,663 &  0.46 & 155.2 : 1 \\
ARP Spoofing    &       5 & 0.0004 & 175,805 : 1 \\
\midrule
\textbf{Total}  & \textbf{1,231,411} & \textbf{100} & N/A \\
\bottomrule
\end{tabular}
\end{table}

\subsection{Statistical Learning-Based Generation}
\label{sec:stat}

The statistical learning pipeline trains 12 class-specific model ensembles, each comprising three components: One-Class SVM (OCSVM), Isolation Forest (IF), and PCA. Table~\ref{tab:sa_params} lists all hyperparameters. Algorithm~\ref{alg:statistical} formalizes the procedure.

\begin{table}[t]
\centering
\caption{Statistical Learning Hyperparameters}
\label{tab:sa_params}
\begin{tabular}{@{}lcc@{}}
\toprule
\textbf{Parameter} & \textbf{Symbol} & \textbf{Value} \\
\midrule
OCSVM regularization & $\nu_c$ & $\min(0.1,\, 1/(n{+}1))$ \\
OCSVM kernel         & $K$     & RBF, $\gamma{=}\text{scale}$ \\
IF contamination     & $c_f$   & $= \nu_c$ \\
IF ensemble size     & $T$     & 100 trees \\
PCA variance retained & $v_c$  & $\min(0.95, \max(0.1, 1{-}10/n))$ \\
PCA weight (hybrid)  & $\alpha$ & 0.7 \\
Gaussian weight      & $1{-}\alpha$ & 0.3 \\
Noise scaling        & $\beta$ & 0.5 \\
Latent noise std     & $\sigma_z$ & 0.1 \\
Target per class     & $M$     & 1,000 packets \\
Max attempts         & $5M$    & 5,000 per class \\
Batch size           & $b$     & 100 \\
\bottomrule
\end{tabular}
\end{table}

\textbf{Adaptive Parameter Selection.} For a class with $n$ training samples, the OCSVM regularization parameter is set as:
\begin{equation}
    \nu_c = \min\left(0.1,\; \frac{1}{n + 1}\right)
\label{eq:nu}
\end{equation}
This formulation prevents the decision boundary from overfitting to sparse classes while retaining sensitivity for well-represented categories. The OCSVM employs an RBF kernel with $\gamma = \texttt{scale}$. Isolation Forest contamination is set identically to Eq.~\eqref{eq:nu}, using 100-tree ensembles (the scikit-learn default) per class. PCA component retention is adapted to class size via $n_{\text{comp}} = \min(0.95, \max(0.1, 1 - 10/n))$, ensuring well-represented classes retain 95\% of explained variance while sparse classes use proportionally fewer components to avoid overfitting.

\textbf{Hybrid Sampling.} Synthetic packets are generated through a two-source mixture:
\begin{equation}
    \hat{x} = \alpha \cdot \text{PCA}^{-1}(z) + (1 - \alpha) \cdot (\boldsymbol{\mu}_c + \boldsymbol{\epsilon} \cdot \boldsymbol{\sigma}_c \cdot \beta)
\label{eq:hybrid}
\end{equation}
where $z \sim \mathcal{N}(\mathbf{0}, 0.1 \cdot \mathbf{I}_p)$ is a latent-space sample with $p$ PCA components, $\boldsymbol{\epsilon} \sim \mathcal{N}(\mathbf{0}, \mathbf{I}_d)$, $\alpha = 0.7$ (PCA weight), and $\beta = 0.5$ (noise scaling factor). The PCA inverse transformation preserves inter-feature correlations critical to IoT traffic patterns, while the direct sampling component introduces controlled variability.

\textbf{Dual-Model Constraint.} A generated packet $\hat{x}$ is accepted only if both models classify it as normal:
\begin{equation}
    \text{accept}(\hat{x}) = \left[\text{OCSVM}(\hat{x}) = +1\right] \wedge \left[\text{IF}(\hat{x}) = +1\right]
\label{eq:dual}
\end{equation}
Rejected packets are discarded and regenerated in batches of 100 until $M = 1{,}000$ valid samples per class are obtained, with a maximum of $5M = 5{,}000$ generation attempts per class to prevent infinite loops on degenerate classes.

\begin{algorithm}[t]
\caption{Statistical Learning Generation}
\label{alg:statistical}
\begin{algorithmic}[1]
\REQUIRE Dataset $\mathcal{D}$, classes $\mathcal{C}$, target count $M=1000$
\ENSURE Synthetic packets $\mathcal{S}_c$ for each $c \in \mathcal{C}$
\FOR{each class $c \in \mathcal{C}$}
    \STATE $\mathcal{D}_c \leftarrow$ samples of class $c$ from $\mathcal{D}$
    \STATE $n \leftarrow |\mathcal{D}_c|$; \quad $\nu_c \leftarrow \min(0.1, 1/(n+1))$
    \STATE Train $\text{OCSVM}_c$ with $\nu_c$, RBF kernel, $\gamma=\text{scale}$
    \STATE Train $\text{IF}_c$ with contamination $= \nu_c$, 100 trees
    \STATE Train $\text{PCA}_c$ retaining $\min(0.95, \max(0.1, 1{-}10/n))$ variance
    \STATE Compute $\boldsymbol{\mu}_c, \boldsymbol{\sigma}_c, \mathbf{l}_c, \mathbf{u}_c$ from $\mathcal{D}_c$
    \STATE $\mathcal{S}_c \leftarrow \emptyset$; \quad $t \leftarrow 0$
    \WHILE{$|\mathcal{S}_c| < M$ \AND $t < 5M$}
        \STATE $z \sim \mathcal{N}(\mathbf{0}, 0.1 \cdot \mathbf{I}_{p})$; \quad $\hat{x}_{\text{pca}} \leftarrow \text{PCA}_c^{-1}(z)$
        \STATE $\boldsymbol{\epsilon} \sim \mathcal{N}(\mathbf{0}, \mathbf{I}_d)$; \quad $\hat{x}_{\text{gauss}} \leftarrow \boldsymbol{\mu}_c + \boldsymbol{\epsilon} \cdot \boldsymbol{\sigma}_c \cdot 0.5$
        \STATE $\hat{x} \leftarrow 0.7 \cdot \hat{x}_{\text{pca}} + 0.3 \cdot \hat{x}_{\text{gauss}}$
        \STATE $\hat{x} \leftarrow \text{clip}(\hat{x}, \mathbf{l}_c, \mathbf{u}_c)$ \hfill $\triangleright$ Boundary enforcement
        \IF{$\text{OCSVM}_c(\hat{x}) = +1$ \AND $\text{IF}_c(\hat{x}) = +1$}
            \STATE $\mathcal{S}_c \leftarrow \mathcal{S}_c \cup \{\hat{x}\}$
        \ENDIF
        \STATE $t \leftarrow t + 1$
    \ENDWHILE
\ENDFOR
\end{algorithmic}
\end{algorithm}

\subsection{Genetic Algorithm-Based Generation}
\label{sec:ga}

The genetic algorithm (GA) treats IoT packet generation as a multi-objective optimization problem. Table~\ref{tab:ga_params} lists all hyperparameters. Algorithm~\ref{alg:ga} provides the pseudocode.

\begin{table}[t]
\centering
\caption{Genetic Algorithm Hyperparameters}
\label{tab:ga_params}
\begin{tabular}{@{}lcc@{}}
\toprule
\textbf{Parameter} & \textbf{Symbol} & \textbf{Value} \\
\midrule
Population size     & $P$       & 200 \\
Elite count         & $e$       & 20 individuals \\
Mutation rate       & $p_m$     & 0.05 \\
Tournament size     & $k$       & 3 \\
Max generations     & $G_{\max}$ & 50 \\
Target fitness      & $f^*$     & 0.9 \\
Stagnation restart  & $s_r$     & 3 generations \\
Noise (seeded init) & $\sigma_s$ & 0.01 \\
Noise (Gaussian init) & $\sigma_g$ & 0.3 \\
Mutation noise      & $\sigma_m$ & $0.1 \cdot r_j$ \\
\bottomrule
\end{tabular}
\end{table}

\textbf{Model Training.} For each class $c$, the GA trains its own OCSVM and IF models using the same adaptive parameterization as the statistical method: $\nu_c = \min(0.1, 1/(n+1))$ for OCSVM with RBF kernel ($\gamma = \text{scale}$), and contamination $= \nu_c$ for IF with 100-tree ensembles. If RBF training fails for a given class, the implementation falls back to a linear kernel with $\nu = 0.01$.

\textbf{Population Initialization.} The initial population of $P = 200$ individuals is constructed from three sources to balance exploitation and exploration:
\begin{itemize}
    \item \textit{Seeded} (25\%, 50 individuals): Real training samples perturbed with $\mathcal{N}(\mathbf{0}, \sigma_s^2 \cdot \mathbf{I})$, $\sigma_s = 0.01$.
    \item \textit{Gaussian} (50\%, 100 individuals): Samples from $\boldsymbol{\mu}_c + \mathcal{N}(\mathbf{0}, \mathbf{I}) \cdot \boldsymbol{\sigma}_c \cdot \sigma_g$, $\sigma_g = 0.3$.
    \item \textit{Uniform} (25\%, 50 individuals): $\mathcal{U}(\mathbf{l}_c, \mathbf{u}_c)$ for full search space coverage.
\end{itemize}

\textbf{Fitness Function.} The fitness of individual $\hat{x}$ balances anomaly model acceptance (80\%) with distributional fidelity (20\%):
\begin{equation}
\small
F(\hat{x}) \!=\! 0.4 \!\cdot\! [\text{OCSVM}(\hat{x})\!=\!+\!1] + 0.4 \!\cdot\! [\text{IF}(\hat{x})\!=\!+\!1] + 0.2 \!\cdot\! e^{-D(\hat{x},c)}
\label{eq:fitness}
\end{equation}
where $D(\hat{x}, c)$ is defined in Eq.~\eqref{eq:fidelity}. The exponential decay rewards proximity to the class centroid while permitting controlled variability necessary for diverse security testing. Maximum achievable fitness is $F = 1.0$ when both models accept the packet and it lies exactly at the class centroid.

\textbf{Selection and Recombination.} Tournament selection ($k=3$) chooses parents. Three crossover operators are applied with uniform probability: (i)~\textit{Uniform crossover}: feature-level random exchange between parents; (ii)~\textit{Single-point crossover}: segment-level recombination at a random feature index; (iii)~\textit{Blend crossover}: $\hat{x}_{\text{child}} = \alpha \cdot \hat{x}_{p_1} + (1 - \alpha) \cdot \hat{x}_{p_2}$, $\alpha \sim \mathcal{U}(0, 1)$.

\textbf{Mutation.} Applied with probability $p_m = 0.05$ per feature via three operators (selected uniformly at random): (i)~\textit{Gaussian}: $\hat{x}_j \leftarrow \hat{x}_j + \mathcal{N}(0, 0.1 \cdot r_j)$ where $r_j = u_j - l_j$ is the feature range; (ii)~\textit{Uniform}: $\hat{x}_j \leftarrow \mathcal{U}(l_j, u_j)$; (iii)~\textit{Boundary reset}: $\hat{x}_j \leftarrow l_j$ or $u_j$ with equal probability. All offspring are clipped to $[\mathbf{l}_c, \mathbf{u}_c]$ post-mutation.

\textbf{Elitism and Restart.} The top $e = 20$ individuals by fitness are preserved across generations. If mean fitness does not improve for $s_r = 3$ consecutive generations, the non-elite portion of the population is re-initialized using the same three-source strategy to escape local optima. The loop terminates when either $M = 1{,}000$ valid unique packets are collected or $G_{\max} = 50$ generations are exhausted.

\begin{algorithm}
\caption{Genetic Algorithm Generation}
\label{alg:ga}
\begin{algorithmic}[1]
\REQUIRE Preprocessed class data $\mathcal{D}_c$, trained $\text{OCSVM}_c$, $\text{IF}_c$
\ENSURE Synthetic packets $\mathcal{S}_c$ with $|\mathcal{S}_c| = M$
\STATE Initialize $\mathcal{P}$ of size $P$ (25\% seeded, 50\% Gaussian, 25\% uniform)
\STATE $\mathcal{S}_c \leftarrow \emptyset$; \quad $g \leftarrow 0$; \quad $\text{stag} \leftarrow 0$
\WHILE{$|\mathcal{S}_c| < M$ \AND $g < G_{\max}$}
    \FOR{each $\hat{x} \in \mathcal{P}$}
        \STATE Compute $F(\hat{x})$ via Eq.~\eqref{eq:fitness}
        \IF{$\text{OCSVM}_c(\hat{x}) = +1$ \AND $\text{IF}_c(\hat{x}) = +1$ \AND unique($\hat{x}$)}
            \STATE $\mathcal{S}_c \leftarrow \mathcal{S}_c \cup \{\hat{x}\}$
        \ENDIF
    \ENDFOR
    \STATE Preserve top $e$ elites by fitness
    \STATE Select parents via tournament ($k = 3$)
    \STATE Apply crossover (uniform / single-point / blend)
    \STATE Apply mutation ($p_m = 0.05$, Gaussian / uniform / boundary)
    \STATE $\hat{x} \leftarrow \text{clip}(\hat{x}, \mathbf{l}_c, \mathbf{u}_c)$ for all offspring
    \STATE $\mathcal{P} \leftarrow$ elites $\cup$ offspring
    \IF{$\bar{F}$ unchanged for $s_r$ generations}
        \STATE Re-initialize non-elite $\mathcal{P}$ \hfill $\triangleright$ Stagnation restart
        \STATE $\text{stag} \leftarrow 0$
    \ENDIF
    \STATE $g \leftarrow g + 1$
\ENDWHILE
\end{algorithmic}
\end{algorithm}

\subsection{Independent Validation Framework}
\label{sec:validation}

The validation framework enforces strict separation between generation and assessment by training fresh OCSVM and IF models exclusively on original ACI IoT 2023 flow-level feature vectors (75 statistical features extracted via CICFlowMeter, encompassing packet length distributions, inter-arrival times, flag counts, and flow duration metrics); no parameters or model weights are shared with generation-phase models.

\textbf{Statistical Learning Validation.} Four independently trained models in two tiers evaluate synthetic packets. \textit{Global models} are trained on a random 10,000-packet sample ($\nu{=}0.01$, contamination${=}0.01$, 100-tree IF ensembles), capturing dataset-wide norms. \textit{Class-specific models} are trained per class using class-aware sampling (up to 5,000 samples) with adaptive parameterization ($\nu_c = \min(0.1, 1/(n{+}1))$); classes with $n < 20$ are excluded from per-class validation. A synthetic set achieves PASS when $A_f < 0.30$ across all four validators.

\textbf{GA Validation.} The GA validator trains independent class-specific OCSVM and IF models with the same adaptive parameterization. A GA synthetic set achieves PASS when $A_f < 0.30$ for both per-class validators.

\section{Experimental Results and Analysis}
\label{sec:results}

This section evaluates both generation methods against the four formal criteria (Section~\ref{sec:metrics}) and the three research questions (Section~I). All experiments use the complete ACI IoT 2023 dataset~\cite{10773916} ($N{=}1{,}231{,}411$ packets, $K{=}12$ attack categories, $d{=}75$ features). Source code is publicly available.\footnote{https://github.com/mayank02raj/Synthetic-Network-Packet-Generation.git}

\subsection{RQ1: Effectiveness of Dual Anomaly-Detection Gating}

Tables~\ref{tab:sa_results} and~\ref{tab:ga_results} report the anomaly rates ($A_f$, Eq.~\eqref{eq:anomaly_rate}) from independently trained validators. Both methods achieved PASS ($A_f < \tau = 0.30$) across all 12 categories under every validation tier, satisfying Criterion~3.

\begin{table}[t]
\centering
\caption{Statistical Learning: Independent Validation Anomaly Rates (\%). Four tiers: global OCSVM/IF (trained on 10,000 cross-class samples, $\nu{=}0.01$, contamination${=}0.01$, 100 trees); class-specific OCSVM/IF (adaptive $\nu_c{=}\min(0.1, 1/(n{+}1))$). Threshold: $\tau{=}0.30$.}
\label{tab:sa_results}
\begin{tabular}{@{}lcccc@{}}
\toprule
\textbf{Attack Category} & \textbf{Global} & \textbf{Global} & \textbf{Class} & \textbf{Class} \\
 & \textbf{OCSVM} & \textbf{IF} & \textbf{OCSVM} & \textbf{IF} \\
\midrule
Benign              & 0.00 & 13.20 & 0.00 & 0.00 \\
ICMP Flood          & 0.00 &  0.00 & 0.00 & 0.00 \\
Slowloris           & 0.00 &  0.00 & 0.00 & 0.00 \\
SYN Flood           & 0.00 &  0.00 & 0.00 & 0.00 \\
UDP Flood           & 0.00 &  0.00 & 0.00 & 0.00 \\
DNS Flood           & 0.00 &  0.00 & 0.00 & 0.00 \\
Dictionary Attack   & 0.00 &  0.00 & 0.00 & 0.00 \\
OS Scan             & 0.00 &  0.00 & 0.80 & 0.00 \\
Port Scan           & 0.00 &  0.00 & 0.00 & 0.00 \\
Ping Sweep          & 0.00 &  0.00 & 0.00 & 0.00 \\
Vulnerability Scan  & 0.00 &  0.00 & 0.00 & 0.00 \\
ARP Spoofing        & 0.00 &  0.00 & 0.00 & 0.00 \\
\midrule
\textbf{Average}    & \textbf{0.00} & \textbf{1.20} & \textbf{0.07} & \textbf{0.00} \\
\bottomrule
\end{tabular}
\end{table}

\begin{table}[t]
\centering
\caption{Genetic Algorithm: Independent Validation Anomaly Rates (\%). Two tiers: class-specific OCSVM and IF (adaptive $\nu_c$, independently trained). Threshold: $\tau{=}0.30$.}
\label{tab:ga_results}
\begin{tabular}{@{}lccc@{}}
\toprule
\textbf{Attack Category} & \textbf{OCSVM (\%)} & \textbf{IF (\%)} & \textbf{Status} \\
\midrule
Benign              & 0.90 & 0.00 & PASS \\
ICMP Flood          & 0.00 & 0.00 & PASS \\
Slowloris           & 0.10 & 0.00 & PASS \\
SYN Flood           & 0.10 & 0.00 & PASS \\
UDP Flood           & 0.00 & 0.00 & PASS \\
DNS Flood           & 0.60 & 0.00 & PASS \\
Dictionary Attack   & 0.00 & 0.20 & PASS \\
OS Scan             & 2.00 & 0.00 & PASS \\
Port Scan           & 1.00 & 0.00 & PASS \\
Ping Sweep          & 0.00 & 0.00 & PASS \\
Vulnerability Scan  & 0.20 & 0.00 & PASS \\
ARP Spoofing        & 2.50 & 0.00 & PASS \\
\midrule
\textbf{Average}    & \textbf{0.62} & \textbf{0.02} & \textbf{PASS} \\
\bottomrule
\end{tabular}
\end{table}

\textbf{Statistical learning.} The four-tier framework returned 0.00\% average on global OCSVM, 1.20\% on global IF, 0.07\% on class-specific OCSVM, and 0.00\% on class-specific IF. The sole elevated rate is 13.20\% for Benign under the global IF tier. The Benign class ($n{=}879{,}027$) exhibits the widest intra-class variance in the dataset, spanning diverse IoT device types and communication patterns. The global IF, trained on a balanced 10,000-packet cross-class sample, captures dataset-wide norms and is thus more sensitive to Benign's distributional breadth. All three remaining tiers returned 0.00\% for Benign, confirming that synthetic Benign packets match the class-specific distribution. The OS Scan class-specific OCSVM rate of 0.80\% similarly reflects tighter class-specific boundaries learned from 15,399 training samples, and remains far below $\tau$.

\textbf{Genetic algorithm.} The two-tier class-specific framework returned 0.62\% average OCSVM and 0.02\% average IF anomaly rates. The GA produces non-zero OCSVM rates across multiple categories (OS Scan 2.00\%, ARP Spoofing 2.50\%, Port Scan 1.00\%, Benign 0.90\%), unlike the statistical method's near-zero uniformity. This difference arises from how each method enforces constraints. The statistical method applies a binary accept/reject gate (Eq.~\eqref{eq:dual}) that discards any packet failing either model. The GA optimizes a composite fitness function (Eq.~\eqref{eq:fitness}) where 80\% of the score derives from anomaly model acceptance and 20\% from distributional fidelity. A packet can achieve high composite fitness with a marginal OCSVM score if compensated by strong IF acceptance and centroid proximity. When evaluated by independently trained validators sharing no parameters with generation-phase models, these marginal packets produce the observed non-zero rates.

\textbf{Validation independence.} The SA validator trains global models on a fresh 10,000-packet sample and class-specific models via class-aware sampling (up to 5,000 per class). The GA validator trains entirely separate per-class models. Both achieve PASS under these independently trained models, confirming that dual gating produces packets generalizing beyond the specific decision boundaries used during synthesis.

\subsection{RQ2: Generation Under Extreme Class Imbalance}

Table~\ref{tab:amplification} reports amplification results across all 12 categories. Both methods generated 1,000 validated packets per class for all categories, satisfying Criterion~4.

\begin{table}[t]
\centering
\caption{Amplification Results: Original Samples ($n$) to 1,000 Synthetic Packets. ``SA Max $A_f$'' and ``GA Max $A_f$'' report the highest single-tier anomaly rate for each method.}
\label{tab:amplification}
\begin{tabular}{@{}lrrcc@{}}
\toprule
\textbf{Attack Category} & $n$ & \textbf{Factor} & \textbf{SA Max} & \textbf{GA Max} \\
 & & & $A_f$ \textbf{(\%)} & $A_f$ \textbf{(\%)} \\
\midrule
ARP Spoofing        &       5 & $200\times$   & 0.00 & 2.50 \\
SYN Flood           &   5,663 & $0.18\times$  & 0.00 & 0.10 \\
UDP Flood           &   6,871 & $0.15\times$  & 0.00 & 0.00 \\
Slowloris           &   7,628 & $0.13\times$  & 0.00 & 0.10 \\
Dictionary Attack   &  12,673 & $0.08\times$  & 0.00 & 0.20 \\
OS Scan             &  15,399 & $0.06\times$  & 0.80 & 2.00 \\
Vulnerability Scan  &  15,915 & $0.06\times$  & 0.00 & 0.20 \\
DNS Flood           &  19,069 & $0.05\times$  & 0.00 & 0.60 \\
Ping Sweep          &  29,433 & $0.03\times$  & 0.00 & 0.00 \\
ICMP Flood          &  91,226 & $0.01\times$  & 0.00 & 0.00 \\
Port Scan           & 179,130 & $0.006\times$ & 0.00 & 1.00 \\
Benign              & 879,027 & $0.001\times$ & 13.20 & 0.90 \\
\bottomrule
\end{tabular}
\end{table}

\textbf{ARP Spoofing ($n = 5$).} This category has a 175,805:1 imbalance ratio against Benign and constitutes the most challenging test case. The statistical method achieved 0.00\% anomaly across all four tiers; the GA achieved 2.50\% OCSVM and 0.00\% IF. Two design choices enable $200\times$ amplification from 5 samples. First, the adaptive regularization $\nu_c = \min(0.1, 1/(n{+}1)) = \min(0.1, 1/6)$, capped at 0.1 (Eq.~\eqref{eq:nu}), prevents the OCSVM decision boundary from collapsing onto sparse training points. Second, the GA's three-source initialization (25\% seeded at $\sigma{=}0.01$, 50\% Gaussian at $\sigma{=}0.3$, 25\% uniform from $[\mathbf{l}_c, \mathbf{u}_c]$) ensures genetic diversity even from 5 seeds. Neither method required fallback mechanisms: the statistical method stayed within its $5M{=}5{,}000$ attempt budget, and the GA converged within $G_{\max}{=}50$ generations.

\textbf{Why standard oversampling fails at $n = 5$.} SMOTE~\cite{chawla2002smote} interpolates between $k$-nearest neighbors, assuming sufficient local density. At $n{=}5$ in $d{=}75$ dimensions, the neighborhood structure is degenerate: any interpolation can produce physically implausible feature combinations (negative packet counts, impossible flag states). Our constraint-enforcing approach avoids this because every packet must pass dual anomaly model acceptance (Eq.~\eqref{eq:dual} or Eq.~\eqref{eq:fitness}) and boundary clamping ($B_c{=}1.0$ via \texttt{np.clip}), regardless of training set size.

\textbf{Scaling behavior across class sizes.} Table~\ref{tab:amplification} reveals no systematic degradation of anomaly rates as $n$ decreases. The GA's highest anomaly rate (2.50\%) occurs at $n{=}5$ (ARP Spoofing), while its second highest (2.00\%) occurs at $n{=}15{,}399$ (OS Scan). For the statistical method, the highest rate (13.20\%) occurs at $n{=}879{,}027$ (Benign) under the global IF tier, while all categories with $n < 100{,}000$ achieved 0.00\% or near-zero rates. This indicates that anomaly rates are driven primarily by the complexity of the class-conditional distribution rather than by training set size, suggesting that the adaptive parameterization (Eq.~\eqref{eq:nu}) successfully compensates for class scarcity.

\subsection{RQ3: Computational Trade-offs}
\label{sec:computational}

Table~\ref{tab:computational} presents the consolidated comparison across computational and quality dimensions.

\begin{table}[t]
\centering
\caption{Consolidated Computational and Quality Comparison. Throughput ratio: statistical $\div$ GA.}
\label{tab:computational}
\begin{tabular}{@{}lcc@{}}
\toprule
\textbf{Metric} & \textbf{Statistical} & \textbf{GA} \\
\midrule
\multicolumn{3}{@{}l}{\textit{Computational Cost}} \\
\quad Total generation time       & $\sim$11\,s    & $\sim$35\,min \\
\quad Throughput (packets/s)      & $\sim$1,091    & $\sim$5.7 \\
\quad Throughput ratio            & \multicolumn{2}{c}{$\sim$190 : 1} \\
\quad Model predictions per packet & 2 (accept/reject) & ${\sim}400$/generation \\
\midrule
\multicolumn{3}{@{}l}{\textit{Quality}} \\
\quad PASS rate (of 12 categories)   & 12/12       & 12/12 \\
\quad Avg.\ anomaly (best tier)      & 0.00\%      & 0.02\% \\
\quad Avg.\ anomaly (worst tier)     & 1.20\%      & 0.62\% \\
\quad Max single-class $A_f$         & 13.20\%     & 2.50\% \\
\quad ARP Spoofing ($n{=}5$) max $A_f$ & 0.00\%    & 2.50\% \\
\quad Boundary compliance $B_c$      & 1.0         & 1.0 \\
\quad Per-class anomaly variance     & Low         & Moderate \\
\quad Validation tiers               & 4           & 2 \\
\bottomrule
\end{tabular}
\end{table}

\textbf{Throughput analysis.} The $\sim$190:1 throughput difference stems from a fundamental architectural distinction. The statistical method performs single-pass sampling with rejection: generate a candidate via hybrid PCA-inverse and Gaussian sampling (Eq.~\eqref{eq:hybrid}), apply boundary clamping, test against dual models, and accept or discard. Each candidate requires exactly two model predictions ($f_{\text{OCSVM}}$ and $f_{\text{IF}}$). The GA evaluates fitness for all $P{=}200$ individuals per generation, performing $2 \times 200 = 400$ model predictions per generation plus tournament selection ($k{=}3$), three crossover operators, three mutation operators, and elitism preservation, across up to $G_{\max}{=}50$ generations per class.

\textbf{Scalability with class complexity.} For the statistical method, generation throughput varies with the complexity of each class's decision boundary rather than with $n$. The dual-model rejection rate determines how many attempts are needed per accepted packet; well-separated classes (e.g., ICMP Flood, Ping Sweep) have high acceptance rates, while the complex Benign class requires more attempts. For the GA, sparse classes like ARP Spoofing ($n{=}5$) require more generations because the fitness landscape is less well-defined with fewer training samples guiding the OCSVM and IF models, while well-represented classes converge more rapidly due to more informative model boundaries.

\textbf{Quality trade-off.} The statistical method's binary gate (Eq.~\eqref{eq:dual}) produces near-uniform anomaly rates (0.00\% under global OCSVM for all 12 classes), indicating high consistency but limited intra-class variability. The GA's composite fitness optimization (Eq.~\eqref{eq:fitness}) produces a wider per-class range (0.00\% to 2.50\%), with a lower worst-tier average (0.62\% vs.\ 1.20\%). This organic variance indicates broader exploration of the valid feature space, which may better serve adversarial robustness evaluation~\cite{9448103, 7467366} and IDS stress testing where traffic diversity within valid boundaries is operationally important.

\textbf{Deployment context.} The throughput and quality profiles suggest distinct operational roles. The statistical method is suited for rapid dataset augmentation where consistency and efficiency are priorities (e.g., overnight IDS training set expansion, real-time security exercise data injection). The GA is suited for scenarios where diversity is the priority (e.g., red team traffic generation, adversarial robustness testing, or generating varied attack representations for comprehensive IDS evaluation).

\subsection{Formal Criterion Satisfaction}

Table~\ref{tab:criteria} maps each formal generation criterion (Section~I) to the empirical evidence demonstrating its satisfaction.

\begin{table}[t]
\centering
\caption{Formal Criterion Verification Summary}
\label{tab:criteria}
\begin{tabular}{@{}p{0.35\columnwidth}p{0.55\columnwidth}@{}}
\toprule
\textbf{Criterion} & \textbf{Evidence} \\
\midrule
1. Generation-time dual gate & SA: binary accept/reject via Eq.~\eqref{eq:dual}; 0.00\% internal anomaly. GA: composite fitness via Eq.~\eqref{eq:fitness}; accepts only $F > 0.8$ \\
2. Boundary compliance $B_c{=}1.0$ & Both methods enforce \texttt{np.clip}$(\hat{x}, \mathbf{l}_c, \mathbf{u}_c)$ prior to acceptance; $B_c{=}1.0$ for all 24,000 packets \\
3. Independent validation $A_f < 0.30$ & SA: max $A_f{=}13.20\%$ (Table~\ref{tab:sa_results}); GA: max $A_f{=}2.50\%$ (Table~\ref{tab:ga_results}); all tiers PASS \\
4. Class coverage (all $K{=}12$) & 1,000 packets per class for all 12 categories including $n{=}5$ ARP Spoofing (Table~\ref{tab:amplification}) \\
\bottomrule
\end{tabular}
\end{table}

All four formal criteria are satisfied by both methods across all 12 attack categories, with no category requiring relaxation of the acceptance threshold or exclusion from evaluation. The independent validation framework, employing models that share no parameters with generation-phase models, provides the objective quality assessment that distinguishes this evaluation from generation-time self-assessment alone.

\subsection{Experimental Hardware and Performance}

All experiments were conducted on an Apple MacBook Pro 16-inch (2024) equipped with an Apple M4 Max chip featuring a 16-core CPU (12 performance cores and 4 efficiency cores), a 40-core GPU, and a 16-core Neural Engine. The system was configured with 128~GB of unified memory and ran macOS Tahoe. Python 3.10 served as the runtime environment with scikit-learn for anomaly detection models and NumPy for numerical computations.

Table~\ref{tab:computational} presents the time cost comparison between the two generation approaches. The Statistical Learning (SA) method completed generation of 12,000 synthetic packets (1,000 per class across 12 attack categories) in approximately 11 seconds, whereas the Genetic Algorithm (GA) approach required approximately 35 minutes for identical output on the same hardware. This represents a speedup factor of approximately 190$\times$ for the SA method over GA.

The substantial difference in computational cost stems from the iterative evolutionary optimization in GA, which requires multiple generations of fitness evaluation, selection, crossover, and mutation operations. In contrast, the SA method leverages pre-trained anomaly detection models for direct sample-and-validate generation, avoiding the iterative population evolution overhead.

\begin{figure}[h]
\centering
\includegraphics[width=\columnwidth, keepaspectratio]{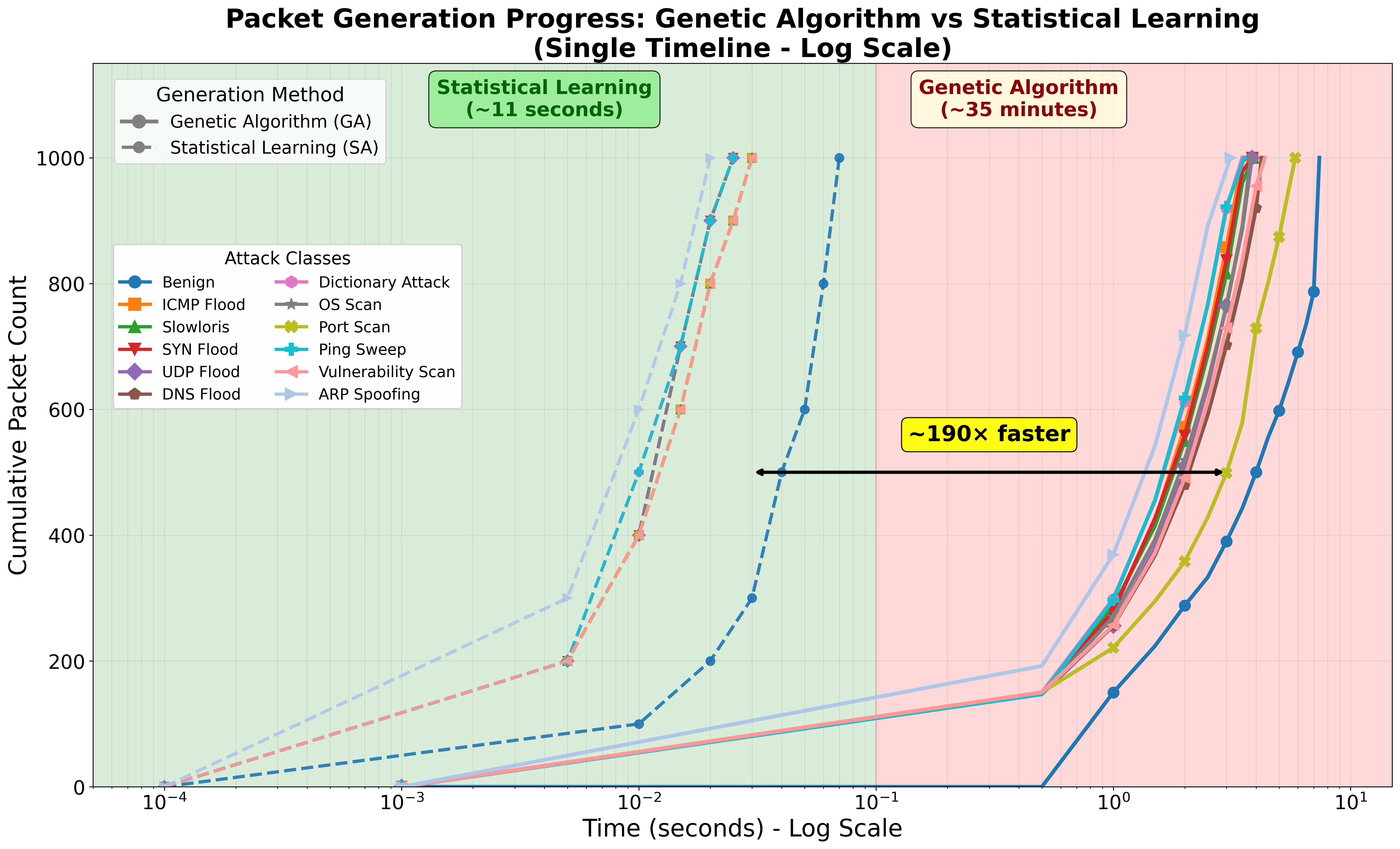}
\caption{Packet generation progress over time for all 12 attack classes. Solid lines represent the Genetic Algorithm (GA) approach, while dashed lines represent the Statistical Learning (SA) approach. The logarithmic time scale reveals that SA completes packet generation approximately 190$\times$ faster than GA, with SA lines clustered in the millisecond range (left) and GA lines extending into the seconds range (right).}
\label{fig:packet_time}
\end{figure}
  The step-wise progression observed in GA (Figure~\ref{fig:packet_time}) reflects the generational evolution process, where plateaus correspond to fitness evaluation and vertical jumps indicate packet acceptance. In contrast, SA exhibits near-linear throughput due to its batch-based sample-and-validate architecture.

\section{Conclusion and Future Research}
\label{sec:conclusion}

This paper presented and compared two constraint-enforcing approaches for generating synthetic IoT network packets: a statistical learning method combining PCA-based latent space sampling with dual OCSVM and Isolation Forest gating, and a genetic algorithm method treating packet generation as a multi-objective optimization problem. Both methods were evaluated on the complete ACI IoT 2023 dataset comprising 1,231,411 packets across 12 attack categories with class imbalance ratios reaching 175,805:1.

The experimental results yield three principal findings. First (RQ1), dual anomaly-detection gating produces synthetic packets that independently trained validators consistently accept: the statistical method achieved 1.20\% average anomaly rate across four validation tiers, and the GA achieved 0.62\% across two tiers, with all 12 categories achieving PASS status under $\tau = 0.30$. Second (RQ2), both methods successfully amplified the 5-sample ARP Spoofing category by $200\times$ to 1,000 validated packets, with maximum anomaly rates of 0.00\% (statistical) and 2.50\% (GA), demonstrating effective operation under extreme class scarcity where standard oversampling methods fail. Third (RQ3), the statistical method offers a $\sim$190:1 throughput advantage ($\sim$1,091 vs.\ $\sim$5.7 packets/s) with near-uniform anomaly rates suited for rapid dataset augmentation, while the GA produces organic per-class variance (0.00\% to 2.50\%) better suited for adversarial robustness evaluation and red team testing.

Several limitations contextualize these findings and motivate future work. Both methods generate individual packets independently without modeling the sequential dependencies inherent in real IoT traffic flows, where device schedules, protocol state machines, and multi-stage attack progressions produce correlated temporal patterns~\cite{9807332, 10.1145/3689935.3690396}. Future work should explore LSTM-based or temporal genetic algorithm approaches for sequence-aware generation that can synthesize realistic multi-step attack campaigns rather than independent packets. Additionally, the current validation framework evaluates statistical properties but does not verify IoT protocol-specific semantics. While boundary clamping ensures no feature exceeds observed ranges, it does not guarantee that resulting feature combinations correspond to valid MQTT, CoAP, Zigbee, or BLE protocol states~\cite{10773916}. Integrating protocol-aware constraint layers would strengthen semantic validity for protocol-level security evaluation.

All experiments use a single dataset (ACI IoT 2023). Although the methodology is dataset-agnostic, requiring only labeled numerical network features as input, empirical validation on additional IoT corpora such as Bot-IoT~\cite{KORONIOTIS2019779}, TON\_IoT, and Edge-IIoTset would strengthen generalizability claims. Finally, the current evaluation measures whether synthetic packets pass anomaly detection validation, but does not assess whether augmenting IDS training sets with these packets improves classifier precision, recall, and F1-score. A downstream evaluation study quantifying the operational impact of constraint-enforcing generation on IDS performance in data-scarce IoT deployments represents an important next step.

\section*{Acknowledgment}
This work was supported by the U.S. Military Academy (USMA) under Cooperative Agreement No. W911NF-22-2-0160. The views and conclusions expressed in this paper are those of the authors and do not reflect the official policy or position of the U.S. Military Academy or U.S. Army.

\begin{IEEEbiography}[{\includegraphics[width=1in,height=1.25in,clip,keepaspectratio]{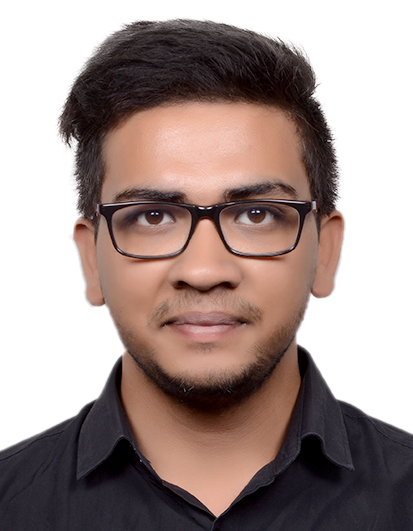}}]
{Mayank Raj}~(Student Member, IEEE \& COMSOC)~ received the M.S.\ degree in Data Science (Thesis Track) with a concentration in cybersecurity from the University of Massachusetts Dartmouth in 2026. His thesis, under \textit{Resilience Engineering of ML-Enabled Open World Recognition for Network Intrusion Detection Systems}, was conducted as a Graduate Research Assistant under Dr.\ Gokhan Kul on a Department of Defense-funded project (Cooperative Agreement No.\ W911NF-22-2-0160) in collaboration with the U.S.\ Military Academy at West Point. He has three years of prior industry experience as a Data Scientist and Software Engineer. His research interests include adversarial machine learning, network intrusion detection, MITRE ATT\&CK-based threat modeling, open-set recognition, and cybersecurity risk assessment. Mr.\ Raj is a member of the IEEE Communications Society, has served as a reviewer for IEEE MILCOM, and as a Technical Program Committee member for the IEEE/IFIP DSN 2026 Workshop. He was a member of UMass Dartmouth's 2026 NCAE Cyber Games team, which placed first nationally.
\end{IEEEbiography}

\begin{IEEEbiography}
[{\includegraphics[width=1in,height=1.25in,clip,keepaspectratio]{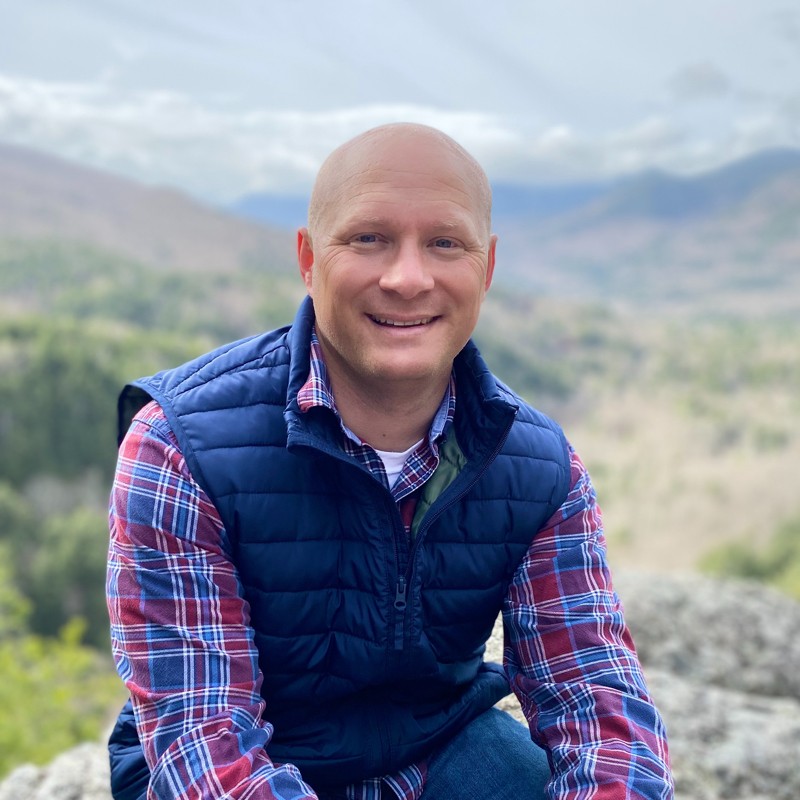}}]{Nathaniel D. Bastian}
(Senior Member, IEEE) received the Ph.D. degree in Industrial Engineering and Operations Research from Pennsylvania State University, University Park, PA, in 2016. He is currently an Assistant Professor in the Department of Electrical Engineering \& Computer Science at the United States Military Academy at West Point, and he serves as Deputy Director of the Robotics Research Center and Principal Investigator of the Laboratory for Artificial Intelligence Research \& Engineering (LAIRE). His primary research interests combine mathematical optimization, decision theory, machine learning, and statistical computing to design and develop secure, robust, and resilient AI-enabled autonomous C5ISRT systems. He has received \$8M in research funding support from DARPA, NSA, OUSD, DEVCOM, AFRL, ONR, etc.
\end{IEEEbiography}

\begin{IEEEbiography}[{\includegraphics[width=1in,height=1.25in,clip,keepaspectratio]{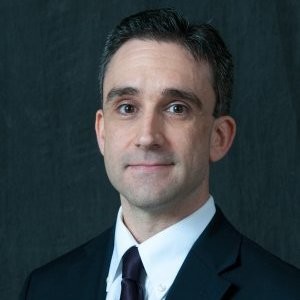}}]{Lance Fiondella}~(Member, IEEE)~received the Ph.D. degree in computer science \& engineering from the University of Connecticut, Storrs, CT, USA, in 2012. He is a professor of Electrical and Computer Engineering and Director of the Cybersecurity Center at the University of Massachusetts Dartmouth, an NSA/DHS designated Center of Academic Excellence in Cyber Research (CAE-R). His research has been funded by the Department of Homeland Security, NASA, the United States Department of Defense, and the National Science Foundation, including a CAREER Award and a CyberCorps Scholarship for Service (SFS) Award. He currently serves as an advisor to Working Group 35 (AI and Autonomous Systems) of the Military Operations Research Society. Previously, he served as associate editor of the Military Operations Research Journal and a technical committee chair of the Annual IEEE Symposium on Technologies for Homeland Security. He also served as the vice-chair of IEEE Standard 1633: Recommended Practice on Software Reliability from 2013-15 and a three-year term as a Member of the Administrative Committee of the IEEE Reliability Society from 2015-2017.
\end{IEEEbiography}

\begin{IEEEbiography}[{\includegraphics[width=1in,height=1.25in,clip,keepaspectratio]{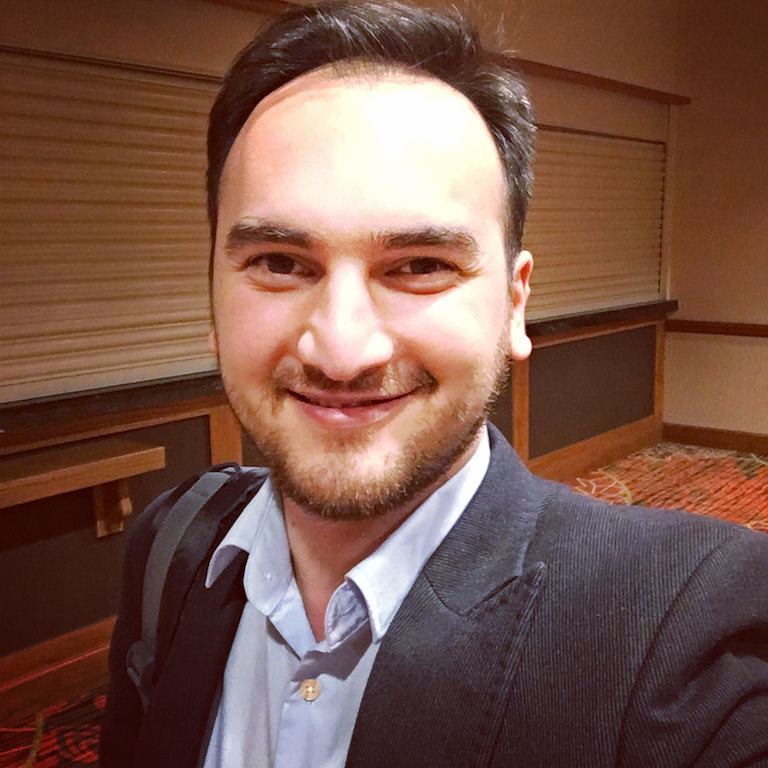}}]{G\"{o}khan Kul}~(Member, IEEE)~received his B.S. degree in Computer Engineering from TOBB University of Economics and Technology, Ankara, T\"{u}rkiye in 2010, his B.A. degree in Business Administration from Anadolu University, Eskisehir, T\"{u}rkiye in 2012, his M.S. degree in Computer Engineering from Middle East Technical University, Ankara, T\"{u}rkiye in 2012, and the Ph.D. degree in Computer Science from the University at Buffalo, SUNY in Buffalo, NY, USA, in 2018. 

He is an Associate Professor at the Department of Computer and Information Science and the Associate Director of the Cybersecurity Center at the University of Massachusetts Dartmouth. Prior to joining UMassD, he was an Assistant Professor at Delaware State University. He has authored or coauthored in reputable journals and conferences on subjects such as intrusion detection, data leakage, concept drift, threat detection, and software vulnerability assessment. His research focuses on AI for cybersecurity. Dr. Kul contributes to research reproducibility efforts as the Co-Chair of VLDB reproducibility program committee.
\end{IEEEbiography}

\end{document}